\documentclass[11pt,twoside]{article}
\begin{document}
\title{\bf Hybrid Quantum Cloning Machine}
\author{S. Adhikari$^{a}$, A. K.
Pati$^{b}$, I. Chakrabarty$^{c}$\thanks{Corresponding
Author:I. Chakrabarty,E-Mail:indranilc@indiainfo.com}, B. S. Choudhury$^{a}$\\
$^{a}$ Bengal Engineering and Science
University, Howrah-711103, West Bengal, India\\
$^{b}$ Institute of Physics, Bhubaneswar-751005, Orissa, India\\
$^{c}$ Heritage Institute of Technology, Kolkata-107, West Bengal,
India}
\maketitle

\begin{abstract}

In this work, we introduce a special kind of quantum cloning
machine called Hybrid quantum cloning machine. The introduced
Hybrid quantum cloning machine or transformation is nothing but a
combination of pre-existing quantum cloning transformations. In
this sense it creates its own identity in the field of quantum
cloners. Hybrid quantum cloning machine can be of two types: (i)
State dependent and (ii) State independent or Universal. We study
here the above two types of Hybrid quantum cloning machines. Later
we will show that the state dependent hybrid quantum-cloning
machine can be applied on only four input states. We will also
find in this paper another asymmetric universal quantum cloning
machine constructed from the combination of optimal universal B-H
quantum cloning machine and universal anti-cloning machine. The
fidelities of the two outputs are different and their values lie
in the neighborhood of $\frac{5}{6} $ .
\end{abstract}
\section{ Introduction }

A fundamental restriction in quantum theory is that quantum
information cannot be copied perfectly $\cite{wz}$ in contrast
with the information we talk about in classical world. Similarly,
it is known that quantum information cannot be deleted against a
copy $\cite{pb,adhi1,adhi2}$. But if we pay some price, then
approximate or exact cloning and deletion operations are possible.
For example, it does not prohibit the possibility of approximate
cloning of an arbitrary state of a quantum mechanical system. The
existence of Universal Copying Machine' (UCM) created a class of
approximate cloning machines which are independent of the
amplitude of the input state $\cite{bh,gm,bbhb}$. The optimality
of such cloning transformations has been verified $\cite{gm}$.
There also exists another class of copying machines which are
state dependent. The original proof of the no-cloning theorem was
based on the linearity of the evolution. Later it was shown that
the unitarity of quantum theory also forbids us from accurate
cloning of non-orthogonal states with certainty $\cite{ay,yuen}$.
But non-orthogonal states secretly chosen from a set can be
faithfully cloned with certain probabilities $\cite{dg,dg1}$ or
can evolve into a linear superposition of multiple-copy states
together with a failure term described by a composite state
$\cite{pati}$ if and only if the states are linearly
independent.\\
The usual scheme of cloning consists of sending a
single photon into an amplifying medium. If there is no photon in
the medium, it spontaneously emit photon of any polarization  but
if the photon is present, the amplifying medium stimulates the
emission of another photon in the same polarization . The quality
of the amplification process is never perfect because spontaneous
emission can never be suppressed $\cite{Scar}$. The
$1\rightarrow2$ quantum cloning machine can be implemented
optically when we take into account the fact that there is a
bridge between stimulated emission and quantum cloning. One of the
first optical experiments using only linear optics by Huang et.al
$\cite{huang}$ that implemented the Buzek-Hillery cloning.\\
Most optical implementations of the $1\rightarrow2$ cloning
machine use parametric down conversion as the amplification
phenomenon. The cloning fidelities obtained in experiments with
parametric
down conversion are $0.81\pm 0.01$ $\cite{Lamas}$ and $0.810\pm 0.08$ $\cite{Martini,Martini1}$.\\

As it is not possible to realize a perfect U-NOT gate which would
flip an arbitrary qubit state, it is necessary to investigate what
is the best approximation to this gate $\cite{Buzek}$. Martini
et.al. reported the experimental realization of universal quantum
machine that performs the best possible approximation to the
universal NOT transformation. The optimal U-NOT transformation
for flipping a single qubit is given by,\\
$U|\psi\rangle_a\otimes|X\rangle_{bc}=
\sqrt{\frac{2}{3}}|\psi\psi\rangle_{ab}|\psi^{+}\rangle_{c}-\sqrt{\frac{1}{3}}(|\psi,\psi^{+}\rangle_{ab}
+|\psi^{+},\psi\rangle_{ab})|\psi\rangle_{c}$\\
where the gate prepared in the state $|X\rangle_{bc}$,
independently of the input state $|\psi\rangle$. The above
transformation describes a process when the original qubit is
encoded in the system 'a', while the flipped qubit is in the
system c. The density operator describing the output state of the
system c is\\
$\rho^{(out)}=\frac{2}{3}|\psi^{+}\rangle\langle \psi^{+}|+\frac{1}{3}|\psi\rangle\langle \psi|$\\
Therefore, the average fidelity of the universal NOT gate is
$F=\langle \psi ^{+}|\rho^{(out)}|\psi ^{+}\rangle=\frac{2}{3}$,
which is exactly same as the fidelity of the optimal state
estimation for single qubit. In the case where a qubit is encoded
into a physical system to utilize the polarization states of the
photon, the U-Not gate can be realized via stimulated emission.
Martini's et.al experiment was based on the proposal that
universal quantum machine such as quantum cloner can be realized
with the help of stimulated emission in parametric down
conversion. The reported experimental fidelity for the optimal
U-NOT
transformation is $0.630\pm 0.008$.\\
In quantum world it is very important to know various limitations
imposed by quantum theory on quantum information. Recently, some
general impossible operations are studied  by Pati $\cite{pati1}$
in detail. This unifies the no-cloning, no-compelementing and
no-conjugating theorems in quantum information theory. Among all
these impossible operations, the impossibility of
'cloning-cum-complementing' quantum machines attracts much
attention here in the sense that it is a combination of cloning
machine and complementing machine where the probabilities of
separately existing cloning machines are $\lambda$ and
$1-\lambda$, respectively. In the same spirit, we can imagine a
hybrid cloning machine which is a superposition of two cloning
machines with appropriate amplitudes $\cite{pati1}$. When the
corresponding probability $\lambda$ takes value between $0$ and
$1$ the resulting combined cloning machines can be identified as a
`Hybrid Cloning Machine' (HCM). Therefore, one can construct
hybrid cloning machine by combining different existing cloning
transformations. Our objective is to study the behavior of such
types of Hybrid cloning machines. Also, we would like to see if
there is any improvements in the fidelity or average fidelity of
cloning under some special combinations. The present work is
organized as follows. In section 2, for the sake of completeness
we recapitulate all the different existing quantum cloning
machines like Wootters-Zurek (WZ) quantum cloning machine,
Buzek-Hillery (BH) quantum cloning machine, Phase Covariant
quantum cloning machine, Pauli Asymmetric quantum cloning machines
and Universal Anti cloning machine. In section 3, we study the
combination of such types of cloning machines which gives  state
dependent hybrid quantum cloning machine. We show here that the
state dependent hybrid quantum cloning machine produces better
quality copy for only four input states. In section 4, we study
the state independent hybrid quantum cloning machines.
Interestingly, we are able to construct here an universal
asymmetric hybrid quantum cloning machine whose fidelity of
copying lie in the neighborhood of the optimal fidelity
$\frac{5}{6}$. Then the conclusion follows.
\section{\bf Descriptions of existing quantum cloning machines}



For the sake of completeness, in this section we briefly discuss
about some existing quantum cloning machines. Then we study the
different combinations of these quantum cloners known as hybrid quantum cloners in the next section.\\
\subsection{The Wootters-Zurek (W-Z) Cloning Machine:}
The Wootters and Zurek (W-Z) quantum cloning machine is a state-
dependent one because it works perfectly for some inputs and badly
for some other. It is defined by the following transformations. In
the computational basis states $|0\rangle$ and $|1\rangle$ it is
given by
\begin{eqnarray}
|0\rangle|Q\rangle\longrightarrow|0\rangle|0\rangle|Q_0\rangle\\
|1\rangle|Q\rangle\longrightarrow|1\rangle|1\rangle|Q_1\rangle.
\end{eqnarray}
Unitarity of the transformation gives
\begin{eqnarray}
\langle Q|Q\rangle=\langle Q_0|Q_0\rangle=\langle
Q_1|Q_1\rangle=1.
\end{eqnarray}
Let us now consider purely  superposition state given by
\begin{eqnarray}
|\chi\rangle=\alpha|0\rangle+\beta|1\rangle.
\end{eqnarray}
For simplicity, we will assume that the probability amplitudes are
real and $\alpha^2+\beta^2=1$.\\
The density matrix of the state $|\chi\rangle$ in the input mode
is given by
\begin{eqnarray}
\rho^{id}=|\chi\rangle\langle
\chi|=\alpha^2|0\rangle\langle0|+\alpha\beta|0\rangle\langle1|+\alpha\beta
|1\rangle\langle0|+\beta^2|1\rangle\langle1|.
\end{eqnarray}
After applying the cloning transformation (1-2) the arbitrary
quantum state (4) takes the form
\begin{eqnarray}
|\psi^{out}\rangle=\alpha|0\rangle|0\rangle|Q_0\rangle+\beta|1\rangle
|1\rangle|Q_1\rangle.
\end{eqnarray}
If it is assumed that two copying machine states $|Q_0\rangle$ and
$|Q_1\rangle$ are orthonormal, then the reduced density operator
$\rho_{ab}^{(out)}$ is given by
\begin{eqnarray}
\rho_{ab}^{(out)}=Tr_x[\rho_{abx}^{(out)}]=\alpha^2|00\rangle\langle00|
+ \beta^2|11\rangle\langle11|.
\end{eqnarray}
The reduced density operators describing the original and the copy
mode are given by
\begin{eqnarray}
\rho_{a}^{(out)}=Tr_b[\rho_{ab}^{(out)}]=\alpha^2|0\rangle\langle0|+
\beta^2|1\rangle\langle1|,\\
\rho_{b}^{(out)}=Tr_a[\rho_{ab}^{(out)}]=\alpha^2|0\rangle\langle0|+
\beta^2|1\rangle\langle1|.
\end{eqnarray}
The copying quality, i.e, the distance between the density matrix
of the input state $\rho_{a}^{(id)}$ and the reduced density
matrices $\rho_{a}^{(out)}$, ($\rho_{b}^{(out)}$) of the output
states can be measured by Hilbert-Schmidt norm. The
Hilbert-Schmidt norm is defined as
\begin{eqnarray}
D_a=Tr[\rho_{a}^{(id)}-\rho_{a}^{(out)}]^2.
\end{eqnarray}
In spite of having other measures of distance between two pure
states Hilbert-Schmidt norm is easier to calculate and also it
serves as a good measure of quantifying the distance between the
pure states. Therefore, we have
\begin{eqnarray}
D_a=2\alpha^2\beta^2=2\alpha^2(1-\alpha^2)
\end{eqnarray}
Since $D_a$ depends on $\alpha^2$, so we have to calculate the
average distortion over all input states, i.e., over all
$\alpha^2$ lying between 0 and 1. Thus, the average distortion is
given by
\begin{eqnarray}
\overline{D_a}=\int^1_0 D_a(\alpha^2)d\alpha^2=\frac{1}{3}.
\end{eqnarray}

\subsection{The Buzek-Hillery (B-H) Cloning Machine}
The Buzek-Hillery cloning machine is a state independent one. This
performs equally well for all input system hence it is a universal
cloner. The BH transformation is given by
\begin{eqnarray}
|0\rangle|Q\rangle\longrightarrow
|0\rangle|0\rangle|Q_0\rangle+[|0\rangle|1\rangle+|1\rangle|0\rangle]
|Y_0\rangle,\\
|1\rangle|Q\rangle\longrightarrow
|1\rangle|1\rangle|Q_1\rangle+[|0\rangle|1\rangle+|1\rangle|0\rangle]
|Y_1\rangle.
\end{eqnarray}
To maintain the unitarity of the transformation, the following
conditions must hold:
\begin{eqnarray}
\langle Q_i|Q_i\rangle + 2\langle Y_i|Y_i\rangle =1,~~~(i=0,1)\\
\langle Y_0|Y_1\rangle=\langle Y_1|Y_0\rangle=0.
\end{eqnarray}
It is further assumed that
\begin{eqnarray}
\langle Q_i|Y_i\rangle=0,~~(i=0,1)\\
\langle Q_0|Q_1\rangle=0.
\end{eqnarray}
The density operator of the output state after copying procedure
is given by
\begin{eqnarray}
\rho_{ab}^{(out)}&=&\alpha^2|00\rangle\langle00|\langle
Q_0|Q_0\rangle+\sqrt{2}\alpha\beta|00\rangle\langle+|\langle
Y_1|Q_0\rangle + \sqrt{2}\alpha\beta|+\rangle\langle00|\langle
Q_0|Y_1\rangle{}\nonumber\\&&+[2\alpha^2\langle
Y_0|Y_0\rangle+2\beta^2\langle Y_1|Y_1\rangle]\langle +|+\rangle
+\sqrt{2}\alpha\beta|+\rangle\langle11|\langle
Q_1|Y_0\rangle{}\nonumber\\&&+\sqrt{2}\alpha\beta|11\rangle\langle+|\langle
Y_0|Q_1\rangle+\beta^2|11\rangle\langle11|\langle Q_1|Q_1\rangle,
\end{eqnarray}
where $|+\rangle=\frac{1}{\sqrt{2}}(|10\rangle+|01\rangle)$. The
reduced density operator describing the original mode can be
obtained by taking partial trace over the copy mode and it reads
as
\begin{eqnarray}
\rho_a^{(out)}= [\alpha^2+\xi(\beta^2-\alpha^2)]
|0\rangle\langle0| + \alpha\beta\gamma |0\rangle\langle1| +
\alpha\beta\gamma |1\rangle\langle0| +
[\beta^2+\xi(\beta^2-\alpha^2)] |1\rangle\langle1|,
\end{eqnarray}
where $\langle Y_0|Y_0\rangle=\langle Y_1|Y_1\rangle\equiv\xi$ and
~~~~~$\langle Y_0|Q_1\rangle=\langle Q_0|Y_1\rangle=\langle
Q_1|Y_0\rangle=\langle Y_1|Q_0\rangle=\frac{\eta}{2}$. The density
operator $ \rho_b^{(out)}$ describing the copy mode is exactly
same as the density operator $\rho_a^{(out)}$ describing the
original mode. Now the Hilbert Schmidt norm for the density
operators $\rho_a^{(id)}$ and $\rho_a^{(out)}$ is given by
\begin{eqnarray}
D_a=2\xi^2(4\alpha^4-4\alpha^2+1)+2\alpha^2\beta^2(\eta-1)^2
\end{eqnarray}
with $0\leq\xi\leq\frac{1}{2}$ and
$0\leq\eta\leq2\sqrt{\xi(1-2\xi)}\leq\frac{1}{\sqrt{2}}$ which
follows from Schwarz inequality.\\
The main criterion in their work was to look out for a copying
machine such that all input states are copied equally well, i.e,
the Hilbert Schmidt norm $D_a$ must be independent of the
parameter $\alpha^2$. Thus, the relation between the parameters
$\xi$ and $\eta$ can be determined from the condition
\begin{eqnarray}
\frac{\delta D_a}{\delta\alpha^2} = 0 \Longrightarrow \eta=1-2\xi.
\end{eqnarray}
Using equation (22), equation (21) reduces to
\begin{eqnarray}
D_a=2\xi^2.
\end{eqnarray}
The value of the parameter $\xi$ can be determined from the second
condition  assumed for the universality criterion of cloning
machine, i.e., the distance between two mode density operators
$\rho_{ab}^{(id)}$ and $\rho_{ab}^{(out)}$ is input state
independent. Mathematically,
\begin{eqnarray}
\frac{\delta D_{ab}^2}{\delta\alpha^2} = 0,
\end{eqnarray}
where $D_{ab}^2=Tr[\rho_{ab}^{(out)}-\rho_{ab}^{(id)}]^2$. Solving
the equation (24) we find $\xi=\frac{1}{6}$. For this value of
$\xi$ the norm $D_{ab}^2$ is independent of $\alpha^2$ and its
value is equal to $\frac{2}{9}$. For $\xi=\frac{1}{6}$, the
deviation of the output from the input is given by
\begin{eqnarray}
D_a=\frac{1}{18}.
\end{eqnarray}

\subsection{ Phase-covariant quantum cloning machine}

Phase covariant quantum cloning machine \cite{bc} can be defined
as
\begin{eqnarray}
|0\rangle |\Sigma\rangle
|Q\rangle|\longrightarrow((\frac{1}{2}+\frac{1}{\sqrt{8}})|0\rangle|0\rangle
+(\frac{1}{2}-\frac{1}{\sqrt{8}})|1\rangle|1\rangle)|\uparrow\rangle+
\frac{1}{2}|+\rangle|\downarrow\rangle,\\
|1\rangle |\Sigma\rangle
|Q\rangle|\longrightarrow((\frac{1}{2}+\frac{1}{\sqrt{8}})|1\rangle|1\rangle
+(\frac{1}{2}-\frac{1}{\sqrt{8}})|0\rangle|0\rangle)|\downarrow\rangle+\frac{1}{2}|+\rangle|\uparrow\rangle.
\end{eqnarray}
The quantum cloning machine defined above can copy the equatorial
states such as $\frac{|0\rangle+e^{i\phi}|1\rangle}{\sqrt{2}}$
with a fidelity $F=\frac{1}{2}+\frac{1}{\sqrt{8}}$ which is
slightly higher than the optimal bound achievable for universal
quantum cloning. The important property of this class that allows
for this higher fidelity is that the coefficients have equal norm.
Due to this property a state dependent term in the final density
matrix of the clones in the cloning transformation becomes
automatically state independent, hence no need for making its
coefficient vanish by tuning the parameters of the cloning
transformation. It had been already shown that if the input state
contains only one unknown parameter, then we are able to construct
a cloning machine which improves the fidelity.\\

\subsection{ Universal asymmetric Pauli cloning machine }

Asymmetric cloning transformation \cite{cerf, cerf1} is given by
\begin{eqnarray}
|0\rangle  |\Sigma\rangle |Q\rangle|\longrightarrow(\frac{1}{\sqrt{1+p^2+q^2}})(|0\rangle|0\rangle|\uparrow\rangle+(p|0\rangle|1\rangle+q|1\rangle|0\rangle)|\downarrow\rangle,\\
|1\rangle |\Sigma\rangle
|Q\rangle|\longrightarrow(\frac{1}{\sqrt{1+p^2+q^2}})(|1\rangle|1\rangle|\downarrow\rangle+(p|1\rangle|0\rangle+q|0\rangle|1\rangle)|\uparrow\rangle.
\end{eqnarray}
Pauli cloning machines (transformations) is nothing but a family
of asymmetric cloning machines that generates two non-identical
approximate  copies of a single quantum bit, each output qubits
emerging from a Pauli channel $\cite{cerf1}$. The asymmetric
quantum cloning machine play an important role in the situation in
which one of the clones need to be a bit better than the
other.\\\\
\begin{tabular}{| c| c| c| c|}
\hline
  parameter (p) & $(F_1)_{PCM}= \frac{(p^2+1)}{2(p^2-p+1)} $ & $(F_2)_{PCM} =\frac{(p^2-2p+2)}{2(p^2-p+1)}$& Difference between
  qualities \\ &  &  &  of the two copies\\ & & &
  $(F_1)_{PCM}\sim (F_2)_{PCM}$\\
  \hline
  0.0 & 0.50 & 1.00 & 0.50 \\
  \hline
  0.1 & 0.55 & 0.99 & 0.44 \\
  \hline
  0.2 & 0.62 & 0.98 & 0.36 \\
  \hline
  0.3 & 0.69 & 0.94 & 0.25 \\
  \hline
  0.4 & 0.76 & 0.89 & 0.13 \\
  \hline
  0.5 & 0.83 & 0.83 & 0.00 (Symmetric copies) \\
  \hline
  0.6 & 0.89 & 0.76 & 0.13 \\
  \hline
  0.7 & 0.94 & 0.69 & 0.25 \\
  \hline
  0.8 & 0.98 & 0.62 & 0.36 \\
  \hline
  0.9 & 0.99 & 0.55 & 0.44 \\
  \hline
  1.0 & 1.00 & 0.50 & 0.50 \\ \hline
\end{tabular}\\\\
The above table represents the quality of the two different
outputs from asymmetric Pauli cloning machine in  terms of the
fidelity for different values of the parameter p. We find that
when $p=0$ or $p=1$, one of the output is totally undisturbed i.e.
contains full information of the quantum state but the other
output contains just 50 percent of the total information. For
$p=0.5$, the Pauli cloning machine reduces to B-H symmetric
quantum cloning machine. We also observe here that the Pauli
quantum cloning machine gives better quality asymmetric copies
when $p=0.4$ and $p=0.6$.

\subsection{Universal anti- cloning machine}
Few years earlier, Gisin and Popescu $\cite{gp}$ discovered an
important fact that quantum information is better stored in two
anti-parallel spins as compared to two parallel spins. This fact
gave birth to a new type of cloning machine called anti-cloning
machine $\cite{sh,gp}$ which generates two outputs, one of the
output has the same direction as the input and the other output
has direction opposite to the input. Song and Hardy $\cite{sh}$
constructed a universal quantum anti-cloner which takes an unknown
quantum state just as in quantum cloner but its output as one with
the same copy while the second one with opposite spin direction to
the input state. For the Bloch vector, an input \textbf{n},
quantum anti-cloner would have the input as
$\frac{1}{2}(\textbf{1}+\textbf{n}.\sigma)$, then it generates two
outputs,$\frac{1}{2}(\textbf{1}+\eta \mathbf{n.\sigma})$ and
$\frac{1}{2}(\textbf{1}-\eta\textbf{n}.\mathbf{\sigma})$, where
$0\leq\eta\leq1$ is the shrinking factor and the fidelity is
defined as $F=\langle\textbf{n}|\rho^{out}|\textbf{n}\rangle=
\frac{1}{2} (1+\eta)$. If spin flipping were allowed then
anti-cloner would have the same fidelity as the regular cloner
since one could clone first then flip the spin of the second copy.
However spin flipping of an unknown state is not allowed in
quantum mechanics. They also showed that the quantum state can be
anti-cloned exactly with non-zero probability.\\
The universal anti-cloning transformation is given by
\begin{eqnarray}
|0\rangle  |\Sigma\rangle  |Q\rangle&\longrightarrow&
\sqrt{\frac{1}{6}}|0\rangle|0\rangle|\uparrow\rangle
+((\frac{1}{\sqrt{2}})e^{icos^{-1}(\frac{1}{\sqrt{3}})}|0\rangle|1\rangle-\frac{1}{\sqrt{6}}|1\rangle|0\rangle)|\rightarrow\rangle+{}\nonumber
\\&&\frac{1}{\sqrt{6}}|1\rangle|1\rangle|\leftarrow\rangle,
\end{eqnarray}

\begin{eqnarray}
|1\rangle  |\Sigma\rangle
|Q\rangle&\longrightarrow&\sqrt{\frac{1}{6}}|1\rangle|1\rangle|\rightarrow\rangle
+((\frac{1}{\sqrt{2}})e^{icos^{-1}(\frac{1}{\sqrt{3}})}|1\rangle|0\rangle-\frac{1}{\sqrt{6}}|0\rangle|1\rangle)|\uparrow\rangle+
{}\nonumber
\\&& \frac{1}{\sqrt{6}}|0\rangle|0\rangle|\downarrow\rangle,
\end{eqnarray}
where$|\uparrow\rangle$,$|\downarrow\rangle$,$|\rightarrow\rangle$,$|\leftarrow\rangle$
are orthogonal machine states. The fidelity of universal
anti-cloner is same as the fidelity of measurement which is equal
to $\frac{2}{3}$ $\cite{mp}$.

\section{State dependent hybrid cloning transformation}


In this section, we study two state dependent cloning machines and
later we find that their average fidelities are greater than the
fidelity of the optimal universal quantum-cloning machine. Since
the quality of the state dependent cloning machine depends on the
input state given to the cloning machine so naturally one may ask
a question why this type of cloning machine is important for
study? Here we give two reasons for this question. First, the
importance of the state dependent cloner lies in the eavesdropping
strategy on some quantum cryptographic system. For example, if the
quantum key distribution protocol is based on two non-orthogonal
states $\cite{bennett}$, the optimal state dependent cloner can
clone the qubit in transit between a sender and a receiver. The
original qubit can then be re-sent to the receiver and the clone
can stay with an eavesdropper who by measuring it can obtain some
information about the bit value encoded in the original. The
eavesdropper may consider storing the clone and delaying the
actual measurement until any further public communication between
the sender and the receiver takes place. This eavesdropping
strategy has been discussed in Ref. $\cite{gh,bruss}$. Second, the
state dependent cloning machines may play an important role when
the cloning machine produces a copy of an arbitrary input state
with better fidelity on average than the optimal universal quantum
cloning machine. Thus an interesting problem would be to construct
a state dependent cloning machine whose average fidelity of
copying is greater than the optimal value $\frac{5}{6}$.\\
{\bf B-H type cloning transformation:} B-H cloning transformation
generally indicates the optimal universal quantum cloning
transformation but in this paper, we relax one condition of
universality of B-H cloning transformation and hence we rename the
B-H cloning transformation as B-H type cloning transformation.
Therefore, although B-H type cloning transformation is
structurally same as the universal B-H cloning transformation but
it is different in the sense that this type of transformation is
state dependent. State dependent ness of the cloning machine
arises because of the relaxation of the condition $\frac{\partial
D_{ab} }{\partial \alpha^2}=0$.\\

\subsection{ Hybridization of two B-H type cloning transformation:}
Here we investigate a new kind of cloning transformation that can
be obtained by combining two different BH type cloning
transformations. This may be given by
\begin{eqnarray}
|\psi\rangle  |\Sigma\rangle
|Q\rangle\otimes|n\rangle\longrightarrow\sqrt{\lambda}[|\psi\rangle|\psi\rangle|Q_{\psi}\rangle
+(|\psi\rangle|\overline{\psi}\rangle+|\overline{\psi}\rangle|\psi\rangle)|Y_{\psi}\rangle]|i\rangle
\nonumber\\+(\sqrt{1-\lambda})[|\psi\rangle|\psi\rangle|{Q_{\psi}^{\prime}}\rangle
+(|\psi\rangle|\overline{\psi}\rangle+|\overline{\psi}\rangle|\psi\rangle)|{Y_{\psi}^{\prime}}\rangle]|j\rangle.
\end{eqnarray}
Unitarity of the transformation gives
\begin{eqnarray}
\lambda(\langle Q_{\psi}|Q_{\psi}\rangle+2\langle
Y_{\psi}|Y_{\psi}\rangle)
+(1-\lambda)(\langle {Q_{\psi}^{\prime}}|{Q_{\psi}^{\prime}}\rangle+2\langle{Y_{\psi}^{\prime}}|{Y_{\psi}^{\prime}}\rangle)=1, \\
2\lambda(\langle Y_{\psi}|Y_{\bar{\psi}}\rangle)
+2(1-\lambda)(\langle{Y_{\psi}^{\prime}}|{Y_{\bar{\psi}}^{\prime}}\rangle)
= 0.
\end{eqnarray}
Equations (33) and (34) is satisfied for all values of
$\lambda(0<\lambda<1)$ if
\begin{eqnarray}
\langle Q_{\psi}|Q_{\psi}\rangle+2\langle
Y_{\psi}|Y_{\psi}\rangle =\langle\acute{Q_{\psi}}|\acute{Q_{\psi}}\rangle+2\langle\acute{Y_{\psi}}|\acute{Y_{\psi}}\rangle =1\\
\langle
Y_{\psi}|Y_{\overline{\psi}}\rangle=\langle\acute{Y_{\psi}}|\acute{Y_{\overline{\psi}}}\rangle=0
\end{eqnarray}
Further we assume that
\begin{eqnarray}
\langle Q_{\psi}|Y_{\psi}\rangle=0=\langle
Q_{\psi}|Q_{\overline{\psi}}\rangle.
\end{eqnarray}
Let $|\chi\rangle=\alpha|0\rangle+\beta|1\rangle$ with
$\alpha^2+\beta^2=1$, be the input state. The cloning
transformation (32) copy the information contained in the input
state $|\chi\rangle$ approximately into two identical states
described by the density operators $\rho_a^{(out)}$ and
$\rho_b^{(out)}$, respectively. The reduced density operator
$\rho_a^{(out)}$ is given by
\begin{eqnarray}
\rho_a^{(out)}&=&|0\rangle\langle0|[\alpha^2+(\beta^2\langle{Y_{1}^{\prime}}|{Y_{1}^{\prime}}\rangle-\alpha^2\langle{Y_{0}^{\prime}}|{Y_{0}^{\prime}}\rangle)+\lambda(\beta^2\langle
Y_1|Y_1\rangle-\alpha^2\langle
Y_0|Y_0\rangle-\beta^2\langle{Y_{1}^{\prime}}|{Y_{1}^{\prime}}\rangle+{}\nonumber\\&&\alpha^2\langle{Y_{0}^{\prime}}|{Y_{0}^{\prime}}\rangle)]
+|0\rangle\langle1|[\alpha\beta(\langle
{Q_{1}^{\prime}}|{Y_{0}^{\prime}}\rangle+\langle
{Y_{1}^{\prime}}|{Q_{0}^{\prime}}\rangle)+{}\nonumber\\&&\lambda\alpha\beta(\langle
Q_{1}|Y_{0}\rangle+\langle Y_{1}|Q_{0}\rangle-\langle
{Q_{1}^{\prime}}|{Y_{0}^{\prime}}\rangle-\langle
{Y_{1}^{\prime}}|{Q_{0}^{\prime}}\rangle)]
+{}\nonumber\\&&|1\rangle\langle0|[\alpha\beta(\langle
{Q_{1}^{\prime}}|{Y_{0}^{\prime}}\rangle+\langle
{Y_{1}^{\prime}}|{Q_{0}^{\prime}}\rangle)+\lambda\alpha\beta(\langle
Q_{1}|Y_{0}\rangle+{}\nonumber\\&&\langle
Y_{1}|Q_{0}\rangle-\langle
{Q_{1}^{\prime}}|{Y_{0}^{\prime}}\rangle-\langle
{Y_{1}^{\prime}}|{Q_{0}^{\prime}}\rangle)]+{}\nonumber\\&&
|1\rangle\langle1|[\beta^2-(\beta^2\langle{Y_{1}^{\prime}}|{Y_{1}^{\prime}}\rangle-\alpha^2\langle{Y_{0}^{\prime}}|{Y_{0}^{\prime}}\rangle)+\lambda(\beta^2\langle
Y_1|Y_1\rangle-\alpha^2\langle
Y_0|Y_0\rangle-{}\nonumber\\&&\beta^2\langle{Y_{1}^{\prime}}|{Y_{1}^{\prime}}\rangle+\alpha^2\langle{Y_{0}^{\prime}}
|{Y_{0}^{\prime}}\rangle)].
\end{eqnarray}
The other output state described by the density operator
$\rho_b^{(out)}$ looks exactly the same as $\rho_a^{(out)}$.\\
Let $\langle Y_0|Y_0\rangle=\langle Y_1|Y_1\rangle=\xi$, $\langle
Q_1|Y_0\rangle=\langle Y_0|Q_1\rangle=\langle
Q_0|Y_1\rangle=\langle Y_1|Q_0\rangle=\frac{\eta}{2}$, \\
$\langle {Y_{0}^{\prime}}|{Y_{0}^{\prime}}\rangle=\langle
{Y_{1}^{\prime}}|{Y_{1}^{\prime}}\rangle={\xi^{\prime}}$ and
$\langle {Q_{1}^{\prime}}|{Y_{0}^{\prime}}\rangle=\langle
{Y_{0}^{\prime}}|{Q_{1}^{\prime}}\rangle=\langle
{Q_{0}^{\prime}}|{Y_{1}^{\prime}}\rangle=\langle
{Y_{1}^{\prime}}|{Q_{0}^{\prime}}\rangle=\frac{{\eta^{\prime}}}{2}$\\
with $0\leq\xi({\xi^{\prime})}\leq1$ and
$0\leq\eta({\eta^{\prime})}\leq2\sqrt{\xi(1-2\xi)}(2\sqrt{{\xi^{\prime}}
(1-2{\xi^{\prime}})})\leq\frac{1}{\sqrt{2}}$.\\
Using above conditions, equation (38) can be rewritten as
\begin{eqnarray}
\rho_a^{(out)}=|0\rangle\langle0|[\alpha^2+{\xi^{\prime}}(\beta^2-\alpha^2)+\lambda(\xi-{\xi^{\prime}})(\beta^2-\alpha^2)]+|0\rangle\langle1|[\alpha\beta({\eta^{\prime}}+\lambda(\eta-{\eta^{\prime}}))]
\nonumber\\+|1\rangle\langle0|[\alpha\beta({\eta^{\prime}}+\lambda(\eta-{\eta^{\prime}}))]+|1\rangle\langle1|[\beta^2-{\xi^{\prime}}(\beta^2-\alpha^2)-\lambda(\xi-{\xi^{\prime}})(\beta^2-\alpha^2)].
\end{eqnarray}
To investigate how well our hybrid cloning machine copy the input
state, we have to calculate the fidelity. Therefore, the fidelity
$F_{HCM}$ is defined by
\begin{eqnarray}
F_{HCM}=\langle\chi|\rho_a^{(out)}|\chi\rangle
=\alpha^4[(1-{\xi^{\prime}})-\lambda(\xi-{\xi^{\prime}})]+\beta^4[(1-{\xi^{\prime}})-\lambda(\xi-{\xi^{\prime}})]
\nonumber\\+2\alpha^2\beta^2[{\xi^{\prime}}+\lambda(\xi-{\xi^{\prime}})+{\eta^{\prime}}+\lambda(\eta-{\eta^{\prime}})].
\end{eqnarray}
Now we get relationship between $\xi
,{\xi^{\prime}},\eta,{\eta^{\prime}}$ by solving the equation
$\frac{\delta F_{HCM}}{\delta \alpha^2}=0$\\
Therefore $\frac{\delta F_{HCM}}{\delta \alpha^2}=0$ implies that
we must have
\begin{eqnarray}
\eta^{\prime}(1-\lambda)+\eta\lambda=1-2\xi^{\prime}-2\lambda(\xi-\xi^{\prime}).
\end{eqnarray}
Using (41), equation (40) reduces to
\begin{eqnarray}
F_{HCM}=(1-\xi^{\prime})-\lambda(\xi-\xi^{\prime}).
\end{eqnarray}
Now the distance between the two mode density operators
$\rho_{ab}^{(out)}$ and
$\rho_{ab}^{(id)}=\rho_a^{(id)}\otimes\rho_b^{(id)}$ is given by
\begin{eqnarray}
D_{ab}=Tr[\rho_{ab}^{(out)}-\rho_{ab}^{(id)}]^2\nonumber\\
=U_{11}^2+2U_{12}^2+2U_{13}^2+U_{22}^2+2U_{23}^2+U_{33}^2,
\end{eqnarray}
where
\begin{eqnarray}
U_{11}&=&\alpha^4-\alpha^2[\lambda(1-2\xi)+(1-\lambda)(1-2{\xi^{\prime}})],
{}\nonumber\\&&
U_{12}=U_{21}=\sqrt{2}\alpha^3\beta-\sqrt{2}\alpha\beta(\eta\frac{\lambda}{2}+(1-\lambda)\frac{{\eta^{\prime}}}{2}){},
\nonumber
\\&& U_{13}=U_{31}=\alpha^2\beta^2{}, \nonumber
\\&&
U_{22}=2\alpha^2\beta^2-(2\xi\lambda+2{\xi^{\prime}}(1-\lambda)){},
\nonumber\\&&
U_{23}=U_{32}=\sqrt{2}\alpha\beta^3-\sqrt{2}\alpha\beta(\eta\frac{\lambda}{2}+(1-\lambda)\frac{{\eta^{\prime}}}{2})
{}, \nonumber\\&&
U_{33}=\beta^4-\beta^2[\lambda(1-2\xi)+(1-\lambda)(1-2{\xi^{\prime}})].
\end{eqnarray}
It is interesting to see that the transformation (32) can behave
as a state dependent cloner if we relax the condition
$\frac{\delta D_{ab}}{\delta \alpha^2}=0$. Therefore, it is
natural to expect that the machine parameters depends on the input
state. Thus, our prime task is to find the relationship between
the machine parameters and the input state that minimizes the
distortion $D_{ab}$. Now we will get an interesting result if we
fix any one of the machine parameters $\xi$ or ${\xi^{\prime}}$ as
$\frac{1}{6}$ . Without any loss of generality we can fix
${\xi^{\prime}}=\frac{1}{6}$ . In doing so, the cloning
transformation (32) reduces to the combination of B-H optimal
universal cloning machine and the B-H type cloning machine.\\
Now, substituting ${\xi^{\prime}}=\frac{1}{6}$ in (44) and using
(41), equation (43) can be rewritten as
\begin{eqnarray}
D_{ab}=V_{11}^2+2V_{12}^2+2V_{13}^2+V_{22}^2+2V_{23}^2+V_{33}^2,
\end{eqnarray}
where
\begin{eqnarray}
 V_{11}&=&\alpha^4-\alpha^2[\lambda(1-2\xi)+(1-\lambda)(\frac{2}{3})]{},
\nonumber\\&&
V_{12}=V_{21}=\sqrt{2}\alpha^3\beta-\sqrt{2}\alpha\beta(\frac{1}{3}-\lambda(\xi-\frac{1}{6}))
{}, \nonumber\\&& V_{13}=V_{31}=\alpha^2\beta^2{},\nonumber\\&&
V_{22}=2\alpha^2\beta^2-(2\xi\lambda+(\frac{1}{3})(1-\lambda)){},
\nonumber\\&&
V_{23}=V_{32}=\sqrt{2}\alpha\beta^3-\sqrt{2}\alpha\beta(\frac{1}{3}-
\lambda(\xi-\frac{1}{6})){}, \nonumber\\&&
V_{33}=\beta^4-\beta^2[\lambda(1-2\xi)+(1-\lambda)(\frac{2}{3})].
\end{eqnarray}
Now we are in a position to determine the relationship between the
machine parameter and the input state that minimizes the
distortion $D_{ab}$. To obtain the minimum value of $D_{ab}$ for
given $\alpha$ and $\lambda $, we solve the equation
\begin{eqnarray}
\frac{\delta D_{ab}}{\delta
\xi}=0\Longrightarrow\xi=\frac{(9\alpha^2\beta^2-2(1-\lambda))}{12\lambda},~~provided~~\lambda\neq0.
\end{eqnarray}
Now, the cloning machine defined by those parameters common to the
whole family of state that one wants to clone. Therefore, it is
clear from equation (47) that the quantum cloning machine can be
applied on the family of states such that
$\alpha^{2}\beta^{2}=constant$. That means the cloning machine can
be applied on just four states
$|\psi^{\pm}\rangle_{1}=\alpha|0\rangle\pm\beta|1\rangle,
|\psi^{\pm}\rangle_{2}=\alpha|1\rangle\pm\beta|0\rangle$.\\
Since the value of the machine parameter $\xi$ cannot be negative,
so the parameter $\lambda$ take values lying in the interval
[$1-\frac{9\alpha^{2}(1-\alpha^{2})}{2}] < \lambda < 1$.\\
Also
\begin{eqnarray}
\frac{\delta^2 D_{ab}}{\delta \xi^2}=16\lambda^2>0.
\end{eqnarray}
Therefore, the equation (47) represents the required relationship
between the machine parameter and the input state which minimizes
$D_{ab}$ and the minimum value of $D_{ab}$ is given by
\begin{eqnarray}
(D_{ab})_{min}=2\alpha^2\beta^2-\frac{9\alpha^4\beta^4}{2}
\end{eqnarray}
which depends on $\alpha^2$ but not on $\lambda$.\\
Substituting $\xi=\frac{(9
\alpha^2(1-\alpha^2)-2(1-\lambda))}{12\lambda}$ and
${\xi^{\prime}}=\frac{1}{6}$ in equation (42), we get\\
$F_{HCM}=1-\frac{3\alpha^2\beta^2}{4}$.\\\\
\begin{tabular}{| c| c| c| c| c|}
\hline Input state $(\alpha^{2})$ & Parameter $\lambda$ & Machine
parameter $(\xi)$ & $(D_{ab})_{min}$ & $F_{HCM}$\\
\hline
0.1~~or~~0.9  & (0.595, 1.0) & (0.0, 0.0675) & 0.14 & 0.93\\
\hline
0.2~~0r~~0.8 & (0.280, 1.0) & (0.0, 0.1200) & 0.21 & 0.88\\
\hline
0.3~~or~~0.7 & (0.055, 1.0) & (0.0, 0.1575) & 0.22 & 0.84\\
\hline
0.4~~or~~0.6 & (0.000, 1.0) & (0.0, 0.1800) & 0.22 & 0.82\\
\hline
0.5 & (0.000, 1.0) & (0.0, 0.1875) & 0.22 & 0.81\\
\hline
\end{tabular}\\\\
The above table shows that there exists several quantum cloning
machines (for different values of $\xi$) which can clone the four
states $\{|\psi^{\pm}\rangle_{1}, |\psi^{\pm}\rangle_{2}\}$ with
the same fidelity. For example, If the input states are chosen
from the set $\{\sqrt{0.1}|0\rangle\pm\sqrt{0.9}|1\rangle,
\sqrt{0.9}|0\rangle\pm\sqrt{0.1}|1\rangle \}$, then corresponding
to different values of the machine parameter $\xi$ $(0 < \xi <
0.0675)$, there exists different quantum cloners which clone the
above states, each with the fidelity 0.93.
\subsection{ Hybridization of B-H type cloning transformation and phase-covariant quantum cloning transformation}
Now, we show that the combination of B-H type cloning
transformation and the phase-covariant quantum cloning
transformation gives a state dependent quantum cloning
transformation which copy the input state having two unknown
parameters with average fidelity greater than
$\frac{1}{2}+\sqrt{\frac{1}{8}}$.\\
The Hybrid cloning transformation is given by
\begin{eqnarray}
|0\rangle |\Sigma\rangle |Q\rangle |n\rangle
\longrightarrow\sqrt{\lambda}[|0\rangle|0\rangle|Q_{0}\rangle
+(|0\rangle|1\rangle+|1\rangle|0\rangle)|Y_{0}\rangle]|i\rangle
\nonumber\\+(\sqrt{1-\lambda})[((\frac{1}{2}+\frac{1}{\sqrt{8}})|0\rangle|0\rangle
+(\frac{1}{2}-\frac{1}{\sqrt{8}})|1\rangle|1\rangle)|\uparrow\rangle+
\frac{1}{2}|+\rangle|\downarrow\rangle)]|j\rangle,\\
|1\rangle |\Sigma\rangle |Q\rangle
|n\rangle\longrightarrow\sqrt{\lambda}[|1\rangle|1\rangle|Q_{1}\rangle
+(|0\rangle|1\rangle+|1\rangle|0\rangle)|Y_{1}\rangle]|i\rangle
\nonumber\\+(\sqrt{1-\lambda})[((\frac{1}{2}+\frac{1}{\sqrt{8}})|1\rangle|1\rangle
+(\frac{1}{2}-\frac{1}{\sqrt{8}})|0\rangle|0\rangle)|\downarrow\rangle+\frac{1}{2}|+\rangle|\uparrow\rangle)]|j\rangle.
\end{eqnarray}
When $\lambda=1$ cloning transformation reduces to B-H type
cloning transformation and when $\lambda=0$ it takes the form of
phase-covariant quantum cloning transformation.\\The cloning
machine (52-53) approximately copy the information of the input
state $|\chi\rangle$ given in (4) into two identical states
described by the reduced density operator
\begin{eqnarray}
\rho=\lambda[(1-\xi)|\chi\rangle\langle\chi|+\xi|\overline{\chi}\rangle\langle\
\overline{\chi}|]+(1-\lambda)[(\frac{1}{2}+\frac{1}{\sqrt{8}})|\chi\rangle\langle\chi|+(\frac{1}{2}-\frac{1}{\sqrt{8}})|\overline{\chi}\rangle\langle\
\overline{\chi}|]
\end{eqnarray}
where $|\bar{\chi}\rangle$ is an orthogonal state to
$|\chi\rangle$. Now, the fidelity is given by
\begin{eqnarray}
F_1=\langle\chi|\rho|\chi\rangle=(\frac{1}{2}+\frac{1}{\sqrt{8}})+\lambda(\frac{1}{2}-\frac{1}{\sqrt{8}}-\xi)
\end{eqnarray}
The hybrid quantum cloning machine constructed by combining the
B-H type cloning transformation and phase-covariant quantum
cloning transformation is state dependent. State dependent ness
condition arises from the fact that B-H type cloning
transformation is state dependent. Consequently, the fidelity
$F_1$ depends on the input state as it depends on the machine
parameter $\xi(\alpha^2)$. We can get the relationship between the
machine parameter $\xi$ associated with the B-H type cloning
machine and the input state $\alpha^2$ by putting $\lambda=1$ in
equation (47). Therefore, the dependence of $\xi$ on $\alpha^2$
can be expressed as
$\xi(\alpha^2)=\frac{3\alpha^2(1-\alpha^2)}{4}$. \\
From the argument given in section (3.1), we find that the hybrid
quantum cloning machine (B-H type cloning transformation + phase
covariant quantum cloning transformation) clone the same four
states $\{|\psi^{\pm}\rangle_{1}, |\psi^{\pm}\rangle_{2}\}$. Also
there is no improvement in the quality of cloning of these four
states. Therefore, this hybrid quantum cloning machine does not
give anything new because it neither involve in cloning of new
family of states nor it gives any improvement in the fidelity of
cloning.
\section{ State independent hybrid cloning transformation}


In this section, we study one symmetric and two asymmetric
universal hybrid quantum cloning machines.

\subsection{ Hybridization of two BH type cloning transformations}
In the preceding section, we find that the quantum cloning machine
obtained by combining two BH type cloning transformations is state
dependent but in this section we will observe that a proper
combination of two BH type cloning transformations can serve as a
state independent cloner also. A hybrid quantum cloning machine
(32) becomes state independent or universal if the fidelity
$F_{HCM}$ and the deviation $D_{ab}$, defined in section 3, both
are state independent. From equation (42), it is clear that
$F_{HCM}$ is state independent. Therefore, the only remaining task
is to show the independence of the deviation $D_{ab}$. We will
find that the deviation $D_{ab}$ is state independent if there
exists a relationship between the parameter $\lambda$ and the
machine parameters $\xi ,{\xi^{\prime}}$. $D_{ab}$ is input state
independent if,
\begin{eqnarray}
\frac{\delta D_{ab}}{\delta \alpha^2}=0
\Longrightarrow[2(\lambda(1-2\xi)+(1-\lambda)(1-2\xi^{\prime}))-3]^2\nonumber\\
-[2(\eta\lambda-(1-\lambda){\eta^\prime})-2]^2+8[2\xi\lambda+2{\xi^\prime}(1-\lambda)]-5=0.
\end{eqnarray}
Using equation (41) in equation (54), we get
\begin{eqnarray}
\lambda=\frac{(6{\xi^{\prime}}-1)}{6({\xi^{\prime}}-\xi)},
\end{eqnarray}
provided $\xi\neq{\xi^{\prime}}$. \\
Using the value of $\lambda$ in (42), we get
\begin{eqnarray}
F_{HCM}=\frac{5}{6}.
\end{eqnarray}
If $\xi={\xi^{\prime}}$ , then there is nothing special about the
transformation (32) because if $\xi={\xi^{\prime}}$ holds then the
transformation (32) simply reduces to B-H cloning machine. The
special feature of the equation (55) is that it makes the
transformation (32) state independent for all values of $\xi$ and
${\xi^{\prime}}$(provided $\xi\neq{\xi^{\prime}}$). This
characteristic of the newly defined cloning machine takes it into
the field of universal cloning machines and creates its
identification as a universal cloner. The introduced universal
cloning machine is optimal also in the sense that the fidelity of
the cloning machine is equal to $\frac{5}{6}$. Although the
machine is universal and optimal for an unknown quantum state but
it is different from B-H cloning machine. It is different in the
sense that B-H cloning machine is state independent for just only
one value of the machine parameter $\xi=\frac{1}{6}$ while the
cloning machine defined by (32) works as a universal cloner for
all values of $\xi$ and ${\xi^{\prime}}$(provided $\xi\neq{\xi^{\prime}}$).\\

\subsection{ Hybridization of optimal universal symmetric B-H cloning transformation and optimal
universal asymmetric Pauli cloning transformation }

Another asymmetric quantum cloning machine can be constructed by
applying hybridization technique. Therefore using the
hybridization procedure we can construct universal asymmetric
quantum cloning machine by combining universal symmetric B-H
cloning transformation and optimal universal asymmetric Pauli
cloning transformation. The Hybrid cloning transformation is given
by
\begin{eqnarray}
&&|0\rangle |\Sigma\rangle |Q\rangle
|n\rangle\longrightarrow\sqrt{1-\lambda}[\sqrt{\frac{2}{3}}|0\rangle|0\rangle
|\uparrow\rangle
+\sqrt{\frac{1}{6}}(|0\rangle|1\rangle+|1\rangle|0\rangle)|\downarrow\rangle]
|i\rangle{}\nonumber\\&&
+\sqrt{\lambda}[(\frac{1}{\sqrt{1+p^2+q^2}})(|0\rangle|0\rangle|\uparrow\rangle
+(p|0\rangle|1\rangle+q|1\rangle|0\rangle)|\downarrow\rangle)]|j\rangle, \\
&&|1\rangle |\Sigma\rangle |Q\rangle
|n\rangle\longrightarrow\sqrt{1-\lambda}[\sqrt{\frac{2}{3}}|1\rangle|1\rangle
|\downarrow\rangle
+\sqrt{\frac{1}{6}}(|0\rangle|1\rangle+|1\rangle|0\rangle)|\uparrow\rangle]
|i\rangle{}\nonumber\\&&
+\sqrt{\lambda}[(\frac{1}{\sqrt{1+p^2+q^2}})(|1\rangle|1\rangle|\downarrow
\rangle+(p|1\rangle|0\rangle+q|0\rangle|1\rangle)|\uparrow\rangle)]|j\rangle,
\end{eqnarray}
\textrm where p + q =1.\\
After taking $|\chi\rangle$ given in (4) as input state by the
cloning machine, the two asymmetric clones emerges as output which
are described by the reduced density operators $\rho_1$ and
$\rho_2$
\begin{eqnarray}
\rho_1=\lambda[(\frac{1}{1+p^2+q^2})((1-q^2+p^2)|\chi\rangle\langle\chi|+q^2I)]+(1-\lambda)[\frac{5}{6}|\chi\rangle\langle\chi|+\frac{1}{6}|\overline{\chi}\rangle\langle\overline{\chi}|], \\
\rho_2=\lambda[(\frac{1}{1+p^2+q^2})((1-p^2+q^2)|\chi\rangle\langle\chi|+p^2I)]+(1-\lambda)[\frac{5}{6}|\chi\rangle\langle\chi|+\frac{1}{6}|\overline{\chi}\rangle\langle\overline{\chi}|].
\end{eqnarray}
Let $F_1$ and $F_2$ denote the fidelities of the two asymmetric
clones.
\begin{eqnarray}
F_1=\frac{5}{6}+(\frac{\lambda}{2})[\frac{(p^2+1)}{(p^2-p+1)}-\frac{5}{3} ~],\\
F_2=\frac{5}{6}+(\frac{\lambda}{2})[\frac{(p^2-2p+2)}{(p^2-p+1)}-\frac{5}{3}~].
\end{eqnarray}
From equation (61) and (62), we can observe that the Hybrid
quantum cloning machine reduces to B-H state independent quantum
cloning machine if $\lambda\rightarrow 0$ and $0\leq p\leq1$ or if
$\lambda\rightarrow 1$ and $p=\frac{1}{2}$.\\ Next our task is to
show that if $F_1>\frac{5}{6}$ then $F_2<\frac{5}{6}$ for all
$\lambda's$ lying between 0 and 1 and vice-versa. Therefore, for
$0<\lambda<1$, we can find $F_1>\frac{5}{6}$ if
$\frac{(p^2+1)}{(p^2-p+1)}>\frac{5}{3}$
\begin{eqnarray}
&&\textrm{i.e. if }   (2p-1)(p-2)<0{}\nonumber\\
&&\textrm{i.e. if }  (2p-1)>0{}\nonumber\\
&&\textrm{i.e. if }  p>\frac{1}{2}. \nonumber
\end{eqnarray}
Now we are going to show that if $p>\frac{1}{2}$ then
$F_2<\frac{5}{6}$. If possible, let $F_2>\frac{5}{6}$ for
$p>\frac{1}{2}$. Therefore, we have
\begin{eqnarray}
F_2>\frac{5}{6}&\Longrightarrow&
\frac{(p^2-2p+2)}{(p^2-p+1)}>\frac{5}{3}{}\nonumber\\
& \Longrightarrow & (2p-1)(p+1)<0{}\nonumber\\
& \Longrightarrow& (2p-1)<0 , \textrm{ Since } p+1>0  {}\nonumber\\
& \Longrightarrow&  p<\frac{1}{2}\nonumber
\end{eqnarray}
which contradicts our assumption. Hence $F_2<\frac{5}{6}$ for
$p>\frac{1}{2}$. Therefore, we can conclude that the fidelities
given in (61) and (62) cannot cross the optimal limit
$\frac{5}{6}$ simultaneously. Next we construct a table below in
which we show that if we made the quality of one of the output
better than the optimal quality then how much far away the quality of the other copy from the optimal one. \\\\
\begin{tabular}{| c| c| c| c| c|}
\hline
  p & $\lambda$ & $F_1=\frac{5}{6}+\frac{\lambda}{2}$ & $F_2=\frac{5}{6}+\frac{\lambda}{2}$& Difference
  \\ & & $[(\frac{p^2+1}{2(p^2-p+1)})-\frac{5}{3}]$& $[(\frac{p^2-2p+2}{p^2-p+1})-\frac{5}{3}]$& between qualities  \\ &&&&of the two copies\\
  \hline
  [0.0,1.0] & 0.0  & 0.83 & 0.83 & 0.00 (symmetric copies)\\
  \hline
  0.0 & [0.1,0.9]&[0.80,0.53] & [0.85,0.98] & [0.05,0.45] \\
  \hline
  0.1 &[0.1,0.9] &[0.81,0.58] & [0.85,0.98] & [0.04,0.40] \\
  \hline
  0.2 & [0.1,0.9]&[0.81,0.64] & [0.85,0.96] & [0.04,0.32] \\
  \hline
  0.3 &[0.1,0.9] &[0.82,0.70] & [0.84,0.93] & [0.02,0.23] \\
  \hline
  0.4 &[0.1,0.9] & [0.83,0.77]& [0.84,0.89] & [0.01,0.12] \\
  \hline
  0.5 &[0.1,0.9] & 0.83 & 0.83 & 0.0 (Symmetric copies) \\
  \hline
  0.6 &[0.1,0.9] &[0.84,0.89] & [0.83,0.77] & [0.01,0.12] \\
  \hline
  0.7 & [0.1,0.9]&[0.84,0.93] & [0.82,0.70] & [0.02,0.23]\\
  \hline
  0.8 & [0.1,0.9]&[0.85,0.96] & [0.81,0.64] & [0.04,0.32] \\
  \hline
  0.9 &[0.1,0.9] &[0.85,0.98] & [0.81,0.58] & [0.04,0.40] \\
  \hline
  [0.0,1.0] & 1.0 & $(F_1)_{PCM}$ & $(F_2)_{PCM}$ &
  $(F_1)_{PCM}\sim(F_2)_{PCM}$\\
  \hline
\end{tabular}\\\\
The above table represents the qualities of the asymmetric copies
of the hybrid cloning machine. We note that the fidelity of the
hybrid quantum cloning machine (B-H cloner + Pauli cloner) depends
on the parameter p and $\lambda$ . From table we observe that one
of the output $(F_1)_{HCM}$   behave as a decreasing function for
$p=0.0$ to $p=0.4$ and for all values of $\lambda$ lying between 0
and 1. At the same time, another output of the asymmetric cloning
machine $(F_2)_{HCM}$ behaves as an increasing function for
$p=0.0$ to $p=0.4$ and for all values of $\lambda$ lying between 0
and 1. The role of the fidelities $(F_1)_{HCM}$ and $(F_2)_{HCM}$
are swapped for $p=0.6$ to $p=0.9$ and for all values of $\lambda$
lying between  0 and 1. Here we observe that the asymmetric hybrid
cloning machine reduces to B-H symmetric cloning machine in two
cases: (i) when $\lambda$ = 0 and $0\leq p\leq1$ and (ii)when
$p=0.5$ and $0.1\leq\lambda\leq0.9$ . Our asymmetric hybrid cloner
also reduces to asymmetric Pauli cloner when $\lambda = 1.0$ and
$0\leq p\leq1$.

\subsection{Hybridization of universal B-H cloning
transformation and universal anti- cloning transformation}

Now we introduce an interesting hybrid quantum-cloning machine,
which is a combination of universal B-H cloning machine and a
universal anti-cloning machine. The introduced cloning machine is
interesting in the sense that it acts like anti-cloning machine.
That means the spin direction of the outputs of the cloner are
antiparallel. We will show later that the newly introduced Hybrid
cloning machine (B-H cloner + Anti-cloner) serve as a better
anti-cloner than the existing quantum anti-cloning machine
$\cite{sh}$. Also we show that if the values of the machine
parameter $\lambda$ is in the neighborhood of 1 then the values of
the two non-identical fidelities lies in the neighborhood of $\frac{5}{6}$ \\
Therefore the introduced anti-cloning transformation is defined by
\begin{eqnarray}
&&|0\rangle |\Sigma\rangle |Q\rangle
|n\rangle\longrightarrow\sqrt{\lambda}[\sqrt{\frac{2}{3}}|0\rangle|0\rangle|\uparrow\rangle
+\sqrt{\frac{1}{6}}(|0\rangle|1\rangle+|1\rangle|0\rangle)|\downarrow\rangle]|i\rangle
+(\sqrt{1-\lambda}){}\nonumber\\&&[\sqrt{\frac{1}{6}}|0\rangle|0\rangle|\uparrow\rangle
+((\frac{1}{\sqrt{2}})e^{icos^{-1}(\frac{1}{\sqrt{3}})}|0\rangle|1\rangle-\frac{1}{\sqrt{6}}|1\rangle|0\rangle)|\rightarrow\rangle+\frac{1}{\sqrt{6}}|1\rangle|1\rangle|\leftarrow\rangle]|j\rangle,\\
&&|1\rangle |\Sigma\rangle |Q\rangle
|n\rangle\longrightarrow\sqrt{\lambda}[\sqrt{\frac{2}{3}}|1\rangle|1\rangle|\downarrow\rangle
+\sqrt{\frac{1}{6}}(|0\rangle|1\rangle+|1\rangle|0\rangle)|\uparrow\rangle]|i\rangle
+(\sqrt{1-\lambda}){}\nonumber\\&&[\sqrt{\frac{1}{6}}|1\rangle|1\rangle|\rightarrow\rangle
+((\frac{1}{\sqrt{2}})e^{icos^{-1}(\frac{1}{\sqrt{3}})}|1\rangle|0\rangle-\frac{1}{\sqrt{6}}|0\rangle|1\rangle)|\uparrow\rangle+\frac{1}{\sqrt{6}}|0\rangle|0\rangle|\downarrow\rangle]|j\rangle,
\end{eqnarray}
where $|\uparrow\rangle$, $|\downarrow\rangle$, $|\rightarrow\rangle$, $|\leftarrow\rangle$ are orthogonal machine states.\\
The above defined cloning machine (63-64) produces two copies of
the input state (4) which are described by the reduced density
operator in mode `a' and mode `b' is given by
\begin{eqnarray}
\rho_a=|0\rangle\langle0|[\lambda(\frac{5\alpha^2}{6}+\frac{\beta^2}{6})+(1-\lambda)(\frac{2\alpha^2}{3}+\frac{\beta^2}{3})]+|0\rangle\langle1|[\lambda\frac{2\alpha\beta}{3}+(1-\lambda)\frac{\alpha\beta}{3}]
\nonumber\\+|1\rangle\langle0|[\lambda\frac{2\alpha\beta}{3}+(1-\lambda)\frac{\alpha\beta}{3}]+|1\rangle\langle1|[\lambda(\frac{5\beta^2}{6}+\frac{\alpha^2}{6})+(1-\lambda)(\frac{\alpha^2}{3}+\frac{2\beta^2}{3})], \\
\rho_b=|0\rangle\langle0|[\lambda(\frac{5\alpha^2}{6}+\frac{\beta^2}{6})+(1-\lambda)(\frac{\alpha^2}{3}+\frac{2\beta^2}{3})]+|0\rangle\langle1|[\lambda\frac{2\alpha\beta}{3}-(1-\lambda)\frac{\alpha\beta}{3}]
\nonumber\\+|1\rangle\langle0|[\lambda\frac{2\alpha\beta}{3}-(1-\lambda)\frac{\alpha\beta}{3}]+|1\rangle\langle1|[\lambda(\frac{5\beta^2}{6}+\frac{\alpha^2}{6})+(1-\lambda)(\frac{2\alpha^2}{3}+\frac{\beta^2}{3})].
\end{eqnarray}
Let $F_a$ and $F_b$ denotes the fidelities of the two copies with
opposite spin direction. Therefore, the fidelities for two outputs
are given by
\begin{eqnarray}
F_a=\frac{5\lambda}{6}+\frac{2(1-\lambda)}{3}, ~
F_b=\frac{5\lambda}{6}+\frac{(1-\lambda)}{3}.
\end{eqnarray}
It is clear from equation (67) that the introduced anti- cloning
machine is asymmetric in nature, i.e., the hybrid quantum cloning
machine resulting from Universal B-H cloning machine and universal
anti-cloning machine behaves as a asymmetric quantum cloning
machine for all values of the parameter $\lambda$ lying between 0
and 1. The two different fidelities given in (67) of the
anti-cloning machine (63-64) can approaches to the optimal value
$\frac{5}{6}$ when the parameter $\lambda$ approaches to one. Here
we should note an important fact that both the fidelities tends to
$\frac{5}{6}$ but not equal to $\frac{5}{6}$ unless $\lambda=1$.
Hence the fidelities $F_a$ and $F_b$ takes different values in the
neighborhood of $\frac{5}{6}$ when the values of $\lambda$ lying
in the neighborhood of 1. For further illustration we construct a table below:\\\\
\begin{tabular}{| c| c| c| c|}
\hline
  parameter $(\lambda)$ & $F_a=\frac{5\lambda}{6}+\frac{2(1-\lambda)}{3}  $ & $F_b=\frac{5\lambda}{6}+\frac{(1-\lambda)}{3}$& Difference between qualities
  \\ & & & of the two copies \\ & & & $F_a\sim F_b$\\
  \hline
  0.0 & 0.67 & 0.33 & 0.34 \\
  \hline
  0.1 & 0.68 & 0.38 & 0.30 \\
  \hline
  0.2 & 0.70 & 0.43 & 0.27 \\
  \hline
  0.3 & 0.72 & 0.48 & 0.24 \\
  \hline
  0.4 & 0.73 & 0.53 & 0.20 \\
  \hline
  0.5 & 0.75 & 0.58 & 0.17 \\
  \hline
  0.6 & 0.77 & 0.63 & 0.14 \\
  \hline
  0.7 & 0.78 & 0.68 & 0.10 \\
  \hline
  0.8 & 0.80 & 0.73 & 0.07 \\
  \hline
  0.9 & 0.82 & 0.78 & 0.04 \\
  \hline
  1.0 & 0.83 & 0.83 & 0.00 (Symmetric copies) \\ \hline
\end{tabular}\\\\
It is clear that both the fidelities of  output copies with
opposite spins are increasing function of the parameter $\lambda$.
Therefore, as $\lambda$ increases, the values of the fidelities
$F_a$ and $F_b$ also increases and approaches towards the optimal
cloning fidelity 0.83. The above Table shows that when $\lambda$ =
0, our Hybrid anti-cloner reduces to anti-cloner introduced by
Song and Hardy $\cite{sh}$. Also when $\lambda=1$ , we observe
that the copies with opposite spin direction changes into the
copies with same spin direction with optimal fidelity.Therefore,
we can conclude that the hybrid anti-cloner performs better than
the existing quantum anti-cloning machine.

\section{ Conclusion}
In this paper we have studied two state dependent hybrid
quantum-cloning machine and three state independent hybrid
quantum-cloning machine. We get few interesting results after
studying the hybrid quantum-cloning machine in detail. First, the
combination of a universal B-H quantum cloning machine and B-H
type quantum cloning machine gives a state dependent hybrid
quantum cloning machine which copy only four input states with
maximum fidelity 0.93. Another hybrid state dependent quantum
cloning machine introduced in this paper is the combination of B-H
type quantum cloning transformation and phase-covariant quantum
cloning transformation. But this type of hybrid quantum cloning
machine does not perform better than other state dependent quantum
cloning machine. Second, the hybridization of two B-H type cloning
transformation also serve as a state independent cloner with
optimal fidelity 5/6 for all values of the machine parameters.
This result is interesting in the sense that the original B-H
quantum cloning machine serve as a universal cloner for just only
one value of the machine parameter but the introduced hybrid
cloner (32) acts as a state independent cloner for all values of
the machine parameters lying in the given range. Third, we
construct here an universal hybrid anti-cloning machine by
combining the universal B-H cloning transformation and universal
anti-cloning transformation. This machine copies an arbitrary
input state with different fidelities of the copies with opposite
spin direction. Although the fidelities are different but the
values of the fidelities lie in the neighborhood of the optimal
value 5/6 provided the machine is constructed in such a way that
the parameter $\lambda$ takes the value close to 1. Thus, our
hybrid anti-cloner can clone an arbitrary input state into two
copies with antiparallel spin direction and improves the quality
of copy upto the optimal quality. Hence collecting all the given
arguments above, we can say that Hybrid quantum cloner performs
better than any other existing individual cloners.\\

\noindent {\bf Acknowledgement:} This work is supported by CSIR
project no.F.No.8/3(38)/2003-EMR-1, New Delhi. S.Adhikari wishes
to acknowledge this support. The work was initiated at the
Institute of Physics (Bhubaneswar) July-August 2005. S.A and I.C
are grateful for the institute's hospitality. S.A and I.C would
like to thank S. Banerjee for her co-operation in completing the
work. B.S.Choudhury acknowledges the support from UGC India under
major research project number,F.8/12/2003 (SR).

\end{document}